\RequirePackage{ifpdf}
\documentclass[12pt,letterpaper]{JHEP3}         
\usepackage{amssymb,amsfonts}

\usepackage{wrapfig}
\usepackage{epsfig}

\def\be{\begin{eqnarray}}
\def\ee{\end{eqnarray}}
\newcommand{\nn}{\nonumber}
\newcommand\para{\paragraph{}}

\newcommand{\eqn}[1]{(\ref{#1})}

\def\Dslash{\,\,{\raise.15ex\hbox{/}\mkern-12mu D}}
\def\Dbarslash{\,\,{\raise.15ex\hbox{/}\mkern-12mu {\bar D}}}
\def\delslash{\,\,{\raise.15ex\hbox{/}\mkern-9mu \partial}}
\def\delbarslash{\,\,{\raise.15ex\hbox{/}\mkern-9mu {\bar\partial}}}
\def\pslash{\,\,{\raise.15ex\hbox{/}\mkern-9mu p}}
\def\calDslash{\,\,{\raise.15ex\hbox{/}\mkern-12mu {\cal D}}}

\newcommand{\RN}{Reissner-Nordstr\"om }

\def\lae{\mathrel{\mathop{\smash{\lower .5 ex \hbox{$\stackrel<\sim$}}}}}
\def\lae{\mathrel{\mathop{\smash{\lower .5 ex \hbox{$\stackrel>\sim$}}}}}

\preprint{DAMTP-2014-89, DCPT-14/71}

\title{Holographic Charge Oscillations}

\author{Mike Blake${}^1$, Aristomenis Donos${}^2$ and David Tong${}^1$\\
${}^1$Department of Applied Mathematics and Theoretical Physics, \\
University of Cambridge, \\ 
Cambridge, CB3 OWA, UK \\
{\tt m.blake, d.tong@damtp.cam.ac.uk}\\
\\ ${}^2$Centre for Particle Theory, Department of Mathematical Sciences, \\
Durham University,  Durham DH1 3LE, UK
\\{\tt aristomenis.donos@durham.ac.uk}}

\abstract{The Reissner-Nordstr\"om black hole provides the prototypical description of a holographic system at finite density. 
We study the response of this system  to the presence of a local, charged impurity. Below a critical temperature, 
the induced charge density, which screens the impurity, exhibits oscillations. 
These oscillations can  be traced to the singularities in the density-density correlation function moving in the complex momentum plane. At finite temperature, the oscillations are very similar to the Friedel oscillations seen in Fermi liquids. However, at zero temperature the oscillations in the black hole background remain exponentially damped, while Friedel oscillations relax to a  power-law. }

\begin{document}
\pagestyle{plain} \setcounter{page}{1}
\newcounter{bean}
\baselineskip16pt \setcounter{section}{0}

\section{Introduction and Summary}

The AdS \RN black hole  offers a holographic description of  a compressible phase of quantum matter at strong coupling \cite{clifford}.  However, despite much study, the nature of this phase remains somewhat mysterious. There is a chemical potential in the boundary theory, yet this results in neither  the condensation of bosons, nor the formation of a sharp Fermi surface.

\para
In recent years, there has been an (almost) exhaustive exploration of the momentum space structure of correlation functions in the \RN black hole and other holographic backgrounds. This has been prompted, in large part, by a desire to understand the fermionic physics captured by these geometries and, not least, how to reconcile the lack of a sharp Fermi surface with the Luttinger theorem 
\cite{frac,huijse,seanhorizon}. At weak coupling, the existence of a Fermi surface means that  low-energy excitations of particles and holes occur at finite momentum. These reveal themselves in  correlation functions as singularities at momentum $k=2k_F$,  twice the Fermi momentum. Holography provides in a framework in which we can ask: what becomes of these $2k_F$ singularities at strong coupling?

\para
 In holographic theories, the spectrum of low-energy, charged excitations can be extracted from  the spectral density; that  is the imaginary, dissipative part of the current-current correlation function. This was studied in some detail in \cite{edalati1,edalati2,shag}. No evidence of singularities in momentum space was seen. Instead, the low-energy physics exhibits  ``local criticality", an unusual form of scale invariance in which time scales but space stands still \cite{hong}. The result is that there are low energy excitations over a range of momenta, $k < \mu_0$, where $\mu_0$ is the chemical potential.

\para
Density-density correlation functions have been  further explored in bulk geometries associated to other locally critical theories \cite{shag,seanpauli}, a number of probe brane models \cite{parny1,parny3,seanpauli} and gapless electron star geometries \cite{lars}. None of these exhibit $2k_F$ singularities. 


\para
There are two exceptions where the bulk, bosonic geometry does exhibit $2k_F$ singularities. The first is somewhat exotic and involves a little string theory (a six-dimensional non-gravitational theory) with a finite density of strings \cite{evajoe}. It seems likely that the physics underlying this behaviour is unrelated to the physics of Fermi surfaces. The second exception occurs in AdS$_3$ where the boundary theory is $d=1+1$ dimensional. Here, the analog of the \RN background is the charged BTZ black hole. It was shown in \cite{faulkneriqbal} that tunnelling events in the bulk, associated to monopoles, give rise to the relevant singular behaviour. Arguments were also presented that suggested such singularities may generically be associated to bulk magnetic degrees of freedom. However, while it is known that monopoles can give rise to a number of interesting effects in higher dimensional holography \cite{bolog,subir,nabil}, the emergence of $2k_F$ singularities does not seem to be among them. At present, it appears  that the beautiful effect described in \cite{faulkneriqbal} is restricted to $d=1+1$ dimensions.

\subsection*{Charge Screening}

A slightly different probe of the momentum structure of the system is offered by its response to a charged impurity. The ground state becomes polarised and the charge is screened. This effect is captured by the static charge susceptibility; that is, the real, reactive part of the density-density correlation function. Although this is also evaluated at frequency $\omega \rightarrow 0$, the charge screening is not necessarily governed by the low-energy excitations of the system. Instead, by the Kramers-Kronig relation, the susceptibility is extracted from  the spectral density by integrating over all frequencies. (This point was emphasised in the holographic context in \cite{seanpauli}). 

\para
Nonetheless, at least for weakly coupled systems, the screening of charge typically is dominated by the lowest energy modes. Indeed, the most visual manifestation of the $2k_F$ singularities of a Fermi surface occur in  {\it Friedel oscillations}. This is the phenomenon in which the induced charge around a
localised impurity varies as $\cos(2k_Fr)$, oscillating between positive and negative.  Further, at zero temperature, this charge decays as a power-law rather than the more typical exponential decay that arises in the Debye or Thomas-Fermi approaches to screening. Heuristically,  the origin of the Friedel oscillations lies in the fact that the lowest energy modes have finite size. These modes enthusiastically cluster around the impurity but, unaware of their own cumbersome nature, over-screen the charge. The story is then repeated, over-exuberance piled upon over-exuberance. The end result is a highly inefficient screening mechanism  and the wonderful rippling patterns of charge that are visible through  scanning tunnelling microscopes.

\para
The purpose of this paper is  to show that similar oscillations in the induced charge occur for the screening of an impurity in the \RN black hole. These oscillations exist despite the fact that there are no sharp features in the spectral density. Moreover, they arise in a rather surprising manner. At high temperatures, $T \gg \mu_0$, there are no oscillations and a localised charge is exponentially screened in the familiar Debye fashion. Correspondingly, the leading poles in the charge susceptibility are on the imaginary momentum axis. As the temperature is lowered, the response changes in a non-analytic manner, reminiscent of a second order phase transition.
 Below a critical temperature, $T_c\sim \mu_0$,  these poles move into the complex momentum plane. This has the effect that, at low temperatures, the induced charge oscillates. The form of these oscillations is similar to the Friedel oscillations of a Fermi liquid. In particular, at finite temperature, both oscillations are exponentially damped.

\para
However, at zero temperature, there is a difference between Friedel oscillations and the charge oscillations we observe in the black hole background. As the temperature approaches zero, we find that many poles coalesce,  resulting in a branch cut in the complex momentum plane. Importantly, this branch cut intersects neither the real nor imaginary axis. Instead it terminates at a position in the complex momentum plane set by the chemical potential and appears to be associated to the AdS$_2$ near-horizon region of the geometry. This is in contrast to a Fermi liquid where the branch cut terminates at real momentum  $k=2k_F$. It means that, at zero temperature, the oscillations around the black hole remain exponentially damped, while the Friedel oscillations become power-law. This also explains why  earlier studies of the spectral density, focussing on real momentum and complex frequency, saw no hint of a sharp momentum structure in the \RN black hole: the momentum structure lies in the complex momentum plane. Such complex momenta are relevant in the study of localised, static perturbations, as opposed to physical, monochromatic excitations which are confined to the real momentum axis.

\para
All the results in this paper can be understood within the regime of linear response. Recently, some very interesting work on impurities in the same theory explored the physics beyond linear response, where an increase in the amplitude of the impurity results in the nucleation of a black hole in the bulk \cite{jorge}.

\section{Charge Screening at Zero Density}

Throughout this paper, we will work with the simplest holographic theory that can describe charge screening; that is, a bulk $d=3+1$ dimensional Einstein-Maxwell theory with action
\be
S_{\mathrm{bulk}}= \int d^4x\sqrt{g}\ \left[ R + \frac{6}{L^2} -\frac{1}{4}F_{\mu \nu}F^{\mu\nu} \right]\label{action}
\ee
Throughout the paper we set the AdS radius to be $L=1$.

\para
The bulk action \eqn{action} provides a holographic description of a boundary conformal field theory in $d=2+1$ dimensions with a conserved $U(1)$ current $J^\mu$ dual to the gauge field $A_{\mu}$.  
This current is associated to a global symmetry on the boundary theory. In any pretence at modelling real materials, one would presumably identify this current with electric charge. The fact that the current is global means that we are neglecting the effect of Coulomb forces between the charges, although the massless excitations of the CFT will mediate other forces.

\para
The neglect of Coulomb forces is a familiar deceit from computations of optical conductivity, both in holography and in more traditional settings,  where it can be justified by the observation that electron-electron interactions are often not the most dominant effect in charge transport. Here, however, we are interested in the screening of charge and, in most materials, the dominant effect {\it is} due to the Coulomb force. This means that we are restricted to situations where the Coulomb force does not hold sway.   Of course, it is always possible to dress our results with Coulomb interactions, a procedure which is typically accomplished using the  random phase approximation. We will not do this in this paper, and the results we present  are for the ``bare" susceptibility\footnote{Our  interest in the screening of a {\it global} charge in the boundary is in contrast to the  screening of non-Abelian $SU(N)$ gauge charge which has been studied  holographically in a number of papers, starting with \cite{andreas}.}.

\para
We start by  describing the screening of charge in the conformal theory with vanishing chemical potential. We will find no surprises, but this gives us the opportunity to recapitulate some basic physics. We will model the impurity as a static chemical potential $\mu(r)$ such that $\mu(r) \rightarrow 0$ as $r \rightarrow 0$. The exact nature of the impurity will not matter too much for us. We only require that  the fall-off is faster than $1/r$, so that this is an irrelevant deformation and does not change the infra-red of the theory. For concreteness, we  choose a simple Gaussian profile 
\be \mu(r) = Ce^{-r^2/2R^2}\label{mur}\ee
However, since we will mostly be interest in the  long-distance physics,  $r \gg R$, our results will not be sensitive to this exact form of the impurity profile.

\para 
Our goal is to compute the response of the charge density $\rho(r) = J^0(r)$ due to the presence of the impurity. Because the profile \eqn{mur} is an irrelevant deformation, we can work perturbatively in the strength of the impurity, given by the dimensionless combination $CR$. We will restrict ourselves to   the regime of linear response. In momentum space, the charge density is controlled by the static susceptibility  $\chi(k)=\langle \rho(k)\rho(-k)\rangle$ and is given by
\be
\rho(k) \sim \chi(k) \mu(k)\nn
\ee
The response function is determined in the limit of frequency $\omega\rightarrow 0$, in which case $\chi(k)$ is real when evaluated on real momenta $k$.  (We will later be interested in the behaviour of the response function for complex $k$).
In the rest of this paper, we will determine how $\chi(k)$ depends on temperature $T$ and chemical potential $\mu_0$.

%
%

\subsubsection*{Zero Temperature}

We start at zero temperature where everything is dictated by conformal invariance. 
On dimensional grounds, in any conformal field theory the charge susceptibility must be given by $\chi(k)\sim k$.

\begin{figure}
  \begin{center}
 \includegraphics[width=2.7in]{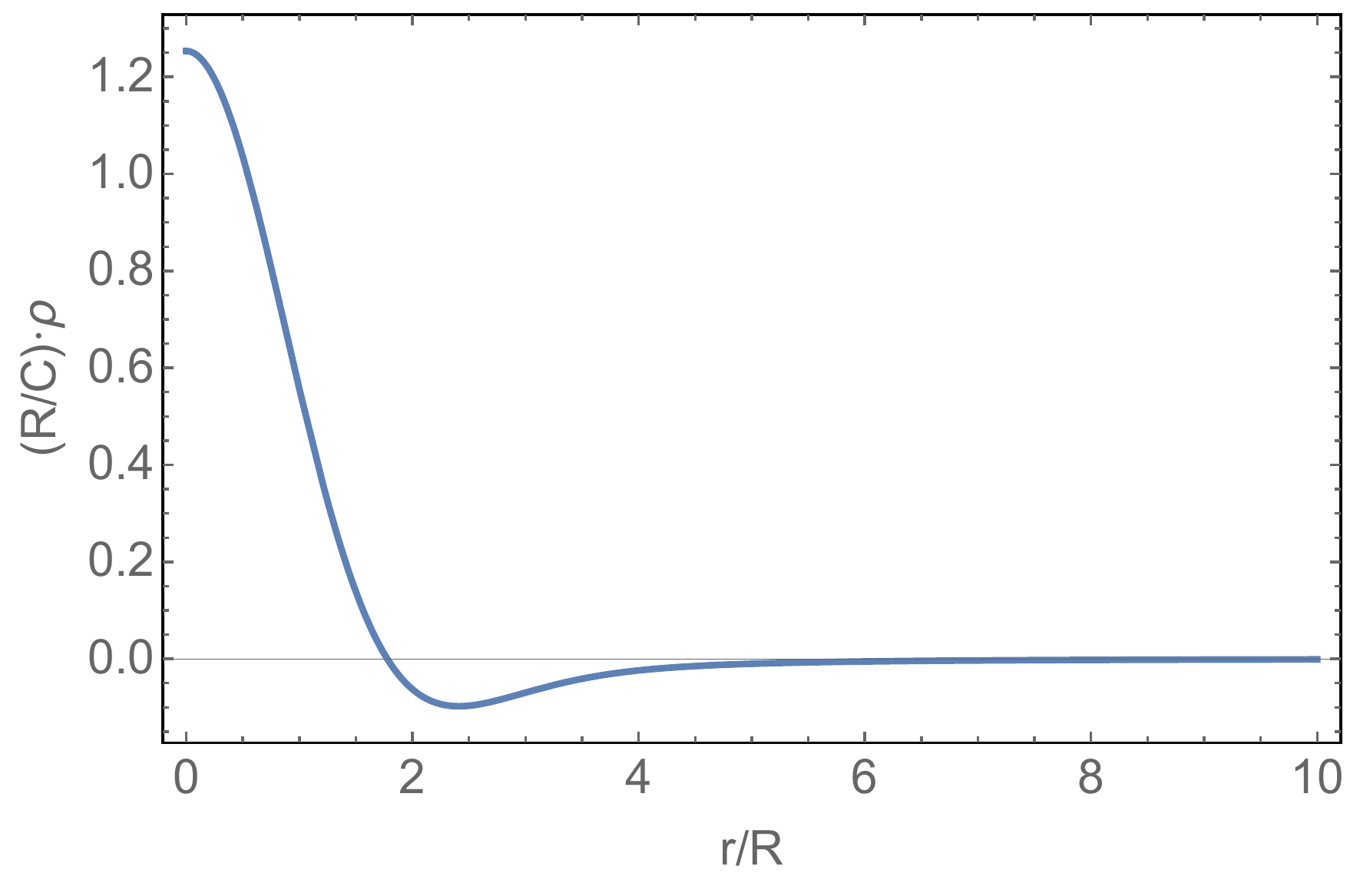}
  \end{center}
\caption{$T=0$, $\mu_0=0$: The charge density  induced by an impurity in the AdS vacuum.}
  \label{figads}
\end{figure}

\para To determine the spatial profile of the induced charge density, we  need only perform the Fourier transform,
\be \rho(r) =  \int \frac{d^2k}{(2\pi)^2}e^{i{\bf k}\cdot {\bf r}}\rho(k)&\sim& CR^2 \int \frac{d^2k}{2\pi}\,\chi(k)\,e^{i{\bf k}\cdot{\bf r}}e^{-R^2k^2/2} \nn\\&=& CR^2\int_0^\infty dk\ k \chi(k) e^{-R^2k^2/2}J_0(kr) 
\label{hankel}\ee
with $J_0(kr)$ a Bessel function. The above expression for the charge density is a Hankel transform. This can be performed analytically  to obtain an exact expression for $\rho(r)$ in terms of Bessel functions. For $\chi(k)=k$, we have
\be
\rho(r) = \frac{C}{2R} \sqrt{\frac{\pi}{2}} e^{-r^2/4 R^2} \bigg[ \frac{r^2}{R^2} I_1(r^2/4R^2) - \bigg(\frac{r^2}{R^2} - 2\bigg) I_0(r^2/4R^2) \bigg]
\nn\ee
The resulting charge density is plotted in Figure \ref{figads}. 

%
%

\para
The key features of the charge density are simple to understand. At large distances, $r\gg R$, the charge falls off as  a power-law, $\rho \sim -CR^2/r^3$, as expected in a scale invariant theory.
 Perhaps the most striking feature is that the induced charge density dips below zero. This too follows from scale invariance, albeit more indirectly. The susceptibility is $\chi(k)\sim k$ which means that the zero-momentum Fourier mode vanishes: $\rho(k=0)=0$. But this, in turn, means that the total, integrated  charge, $Q=\int d^2r\, \rho(r)$, vanishes. The induced charge must therefore dip below zero somewhere. 

\subsubsection*{Finite temperature}

We now turn on finite temperature, $T\neq 0$. To study this situation, we turn to the gravitational description of the ground state given by the AdS Schwarzchild black hole, with metric
\be ds^2 = \frac{1}{z^2}\left(-f(z)dt^2 + \frac{dz^2}{f(z)} + dx^2 + dy^2 \right)\label{metric}\ee
where
\be f(z) = 1 - \left(\frac{z}{z_+}\right)^3\nn\ee
Here  $z$ is the radial, bulk coordinate such that the boundary lies at $z=0$. The black hole  describes the boundary theory at temperature $T=3/4\pi z_+$. 

\para
We perturb the background with the localised boundary chemical potential \eqn{mur}. This is, by now, the kind of calculation that is holographic bread and butter. (For a review of holographic linear response theory, see, for example, \cite{seanlecture}). In this simple case, we need only consider the Maxwell equation for the temporal gauge field, $\delta A_t(z, \vec{x}) = \delta A_t(z) e^{i \vec{k}. \vec{x}}$, which reads
\be
f(z) \delta A_t'' - k^2 \delta A_t = 0
\ee
We solve this equation numerically, subject to regularity at the horizon and extract the susceptibility as the ratio $\chi(k) = -\delta A_t'/\delta A_t$  as $z \rightarrow 0$. 

\begin{figure}
\label{figbh}
\begin{center}
\includegraphics[width=2.6in]{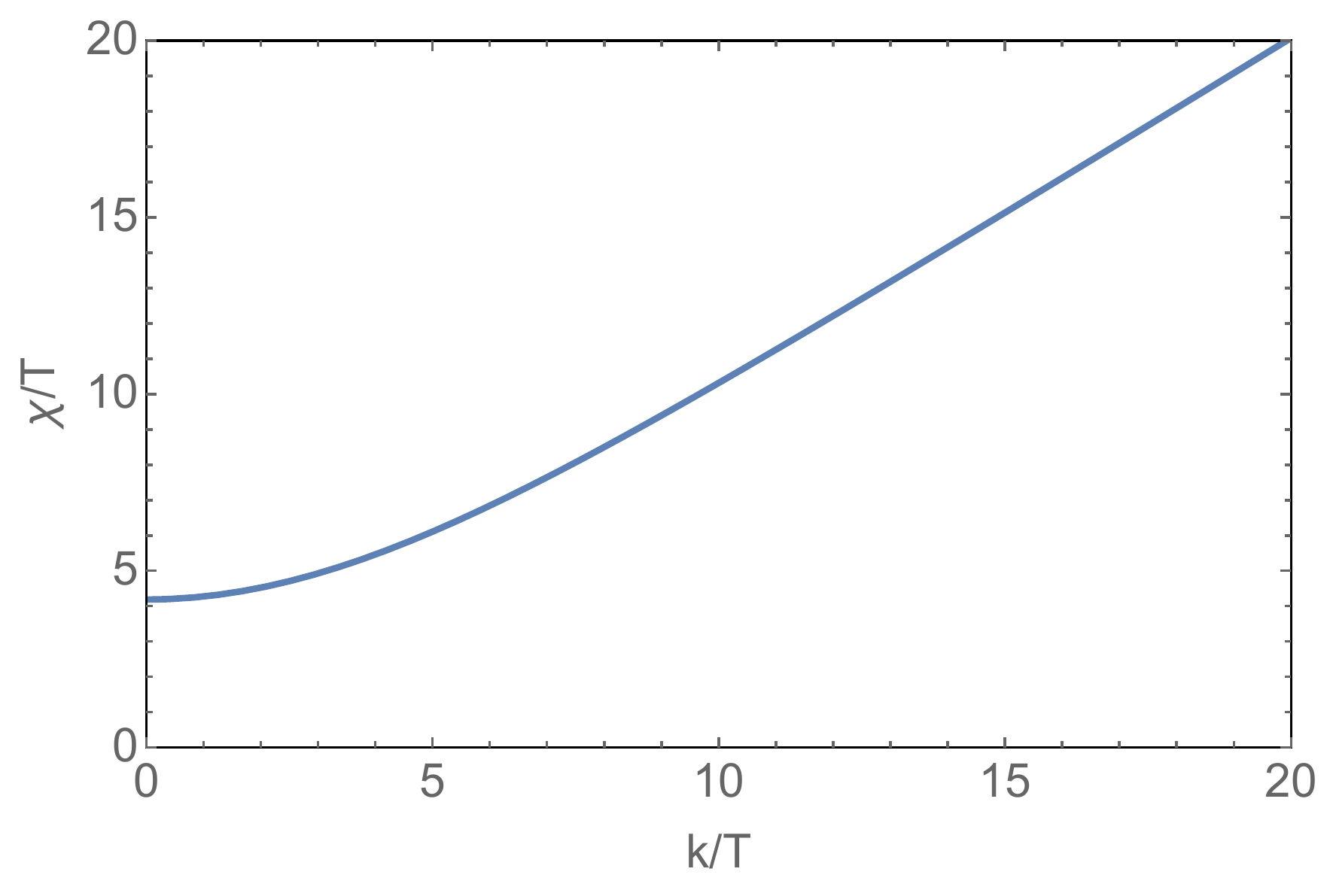}\ \ \ \ \ \ \ \ 
\includegraphics[width=2.6in]{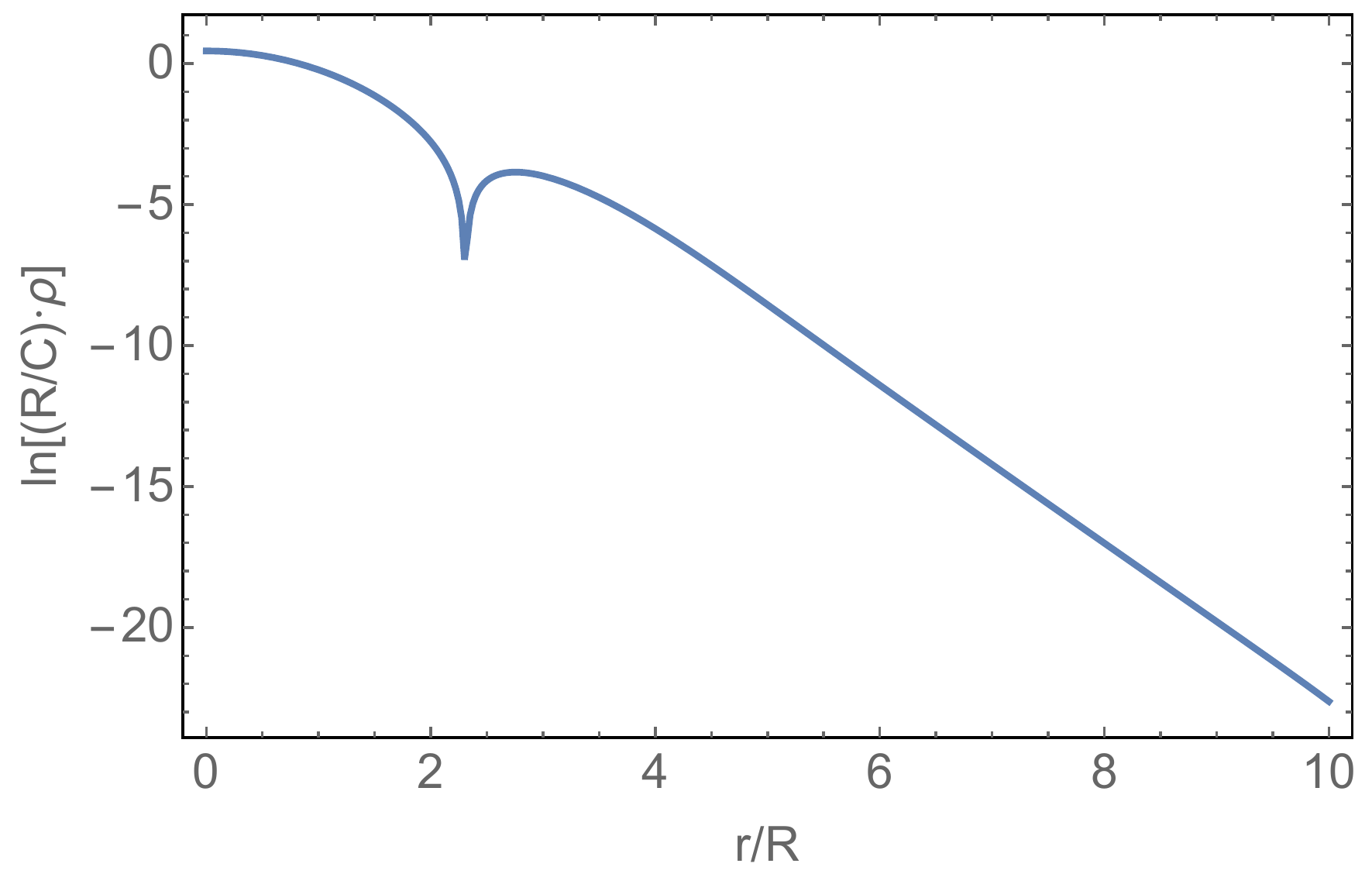}
\caption{$T\neq 0$, $\mu_0=0$: The susceptibility $\chi(k)$ on the left. The finite temperature response, $\log(|R\rho/C|)$, is shown on the right for an impurity of width $RT=3/4\pi$.}
\end{center}
\end{figure}

\para
The susceptibility is plotted on the left of  Figure \ref{figbh} as a function of $k/T$. At high momenta, $k\gg T$ it passes over to the scale invariant result $\chi(k) \sim k$ as expected. However, it deviates from this result at low momenta  and, in  particular, $\chi(k=0)\neq 0$. Correspondingly, the integrated charge is now non-vanishing. This is seen in the right of Figure \ref{figbh} where, because the induced charge density now decays exponentially, we have plotted $\log(|R\rho/C|)$ on the vertical axis. Of course, in a log-plot we must first take the absolute value of $\rho$ which means that sign of the induced charge is no longer obvious. Instead, the cusp  reveals where $\rho$ passes through zero. (In an analytic treatment, the cusp itself would reach down to negative infinity).

\begin{figure}
\begin{center}
\includegraphics[width=2.3in]{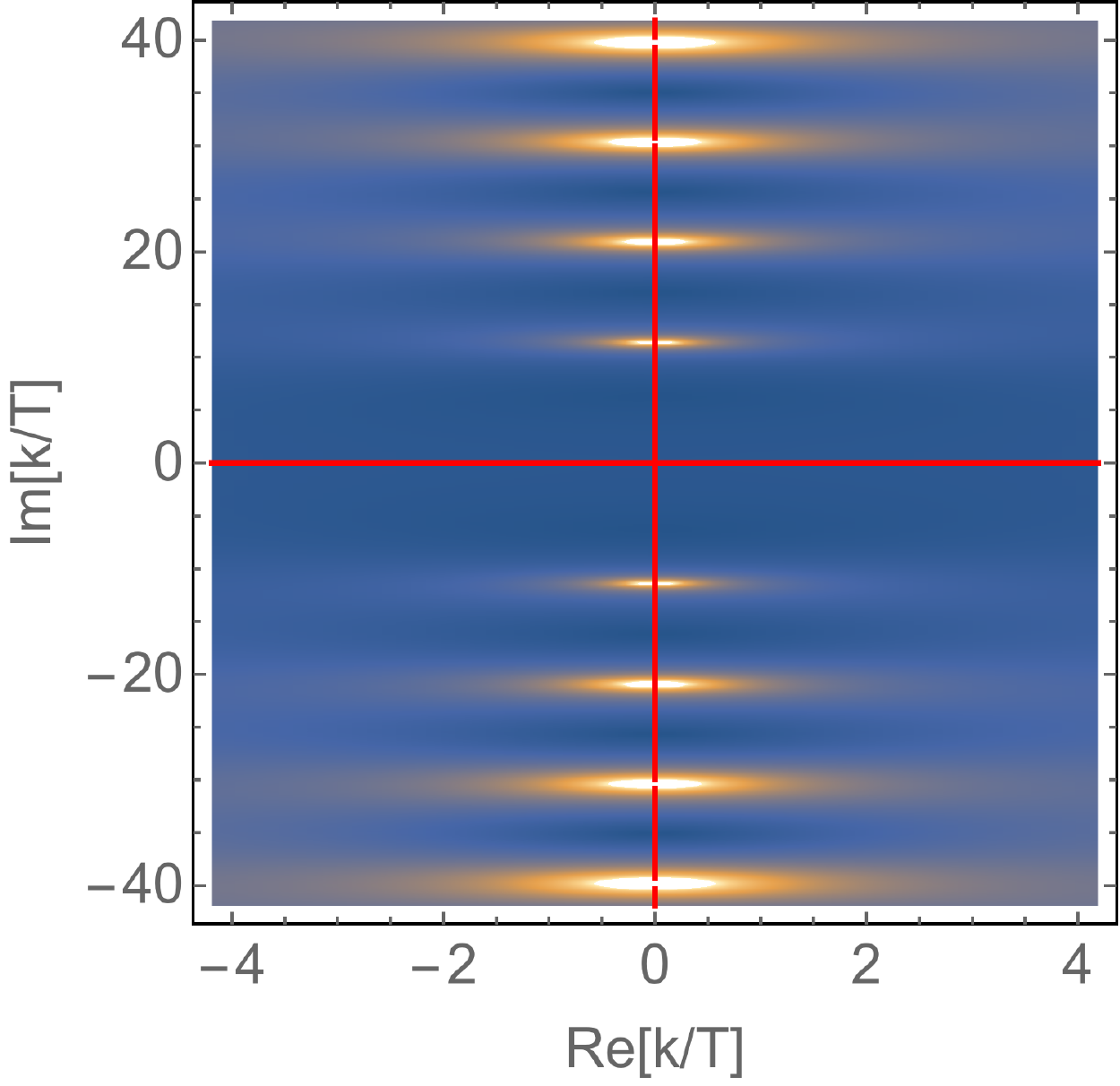}
\caption{$T\neq 0$, $\mu_0=0$: The absolute value of $\langle \rho(k)\rho(-k)\rangle$ correlator in the complex $k_x$-plane. The bright spots are the poles.}
\label{figpoles}
\end{center}
\end{figure}

\para
At large distances, $r\gg R$ and $T^{-1}\gg R$, the charge density  decays exponentially, rather than the power-law that we saw at $T=0$. In fact, one can check that the asymptotic induced charge takes the form
\be \rho(r)\sim CR^2\, \frac{e^{-r/\lambda}}{\sqrt{r\lambda^5}}\label{goeslikethis}\ee
where the screening length scales as $\lambda \sim 1/T$. This is the normal Debye screening behaviour, expected of any relativistic quantum field theory at finite temperature.
\para
The  asymptotic form \eqn{goeslikethis}  can be seen in the complex momentum plane. We align the momentum along the $x$-direction and evaluate $\chi(k_x)$. 
The  density-density correlation function exhibits a series of poles that lie strictly on the imaginary momentum axis, the first of which occurs at $k=i/\lambda \sim iT$. This string of poles is shown in Figure \ref{figpoles}. At large distances, these poles dominate  the Hankel transform of the susceptibility, giving rise to the exponential behaviour \eqn{goeslikethis}.

\para
Note that there is no analyticity in complex momentum: in contrast to complex frequency, poles appear in both the upper and lower halves of the $k_x$-plane. This reflects the fact that complex momentum is not relevant in  a translationally invariant system since the resulting response will be of the form $e^{ikx}$ which, for complex $k$, grows in either the positive or negative $x$-direction. However, as is familiar from quantum mechanics, the presence of a delta-function impurity allows us to pick up contributions from the upper-half plane for $x>0$ and the lower-half plane for $x<0$,  with the impurity providing  the necessary discontinuity in the derivative of the response at the origin.

\para
The static screening properties of the AdS Schwarzchild black hole in AdS$_5$ were previously determined in \cite{hou,liu} with broadly similar results. The charge susceptibility was also  compared with weakly coupled ${\cal N}=4$ super Yang-Mills. Curiously, it was claimed  that the screening is weaker at strong coupling and stronger at weak coupling. (For example, the Debye screening length $\lambda$ was argued to increase as the 't Hooft coupling is increased).

\section{Charge Screening at Finite Density}

We now turn to our main interest: screening at finite density. We turn on a constant chemical potential, $\mu_0$, on the boundary theory. The resulting bulk geometry is the \RN black hole. The metric again takes the form \eqn{metric},  with 
\be f(z) = 1 - \left(1+\frac{z_+^2 \mu_0^2}{4}\right)\left(\frac{z}{z_+}\right)^3 + \frac{z_+^2\mu_0^2}{4}\left(\frac{z}{z_+}\right)^4\nn\ee
This is accompanied by a profile for the Maxwell field,
\be A_0 = \mu_0\left(1-\frac{z}{z_+}\right)\nn\ee
In this geometry, the boundary CFT field theory is warmed to  temperature
\be T = \frac{1}{4\pi z_+}\left(3 - \frac{z_+^2\mu_0^2}{4}\right)\nn\ee
We once again perturb the boundary theory by a localised, charged impurity. The chemical potential is taken to be
\be \mu(r) = \mu_0 + Ce^{-r^2/2R^2}\nn\ee
As in the previous section, the linear response is computed via the susceptibility $\chi(k) =\langle\rho(k)\rho(-k)\rangle$.

\para
There is one aspect of the static susceptibility which is more easily computed with the introduction of $\mu$. This is the zero mode $\chi(k=0)$. As we saw above, this controls the integrated induced charge due to an impurity. It is given analytically by
\be
\chi(k=0)=\frac{\partial Q\left(T,\mu\right)}{\partial \mu}=\frac{2\,\pi\,T}{3}+\frac{1}{3}\,\frac{3\,\mu^{2}+8\,\pi^{2}\,T^{2}}{\sqrt{3\,\mu^{2}+16\,\pi^{2}\,T^{2}}}.
\ee
In  the limit $T\gg\mu$,  we have  $\chi(k=0) =4\pi T/3$. This agrees with the $k=0$ limit of the susceptibility shown on the left  of Figure \ref{figbh}.

\para
To go beyond the $k=0$ limit of the susceptibility we must work numerically. This  
is somewhat more involved than for the Schwarzchild black hole because the gauge field perturbation $\delta A_t$ now back-reacts on the metric. Working in radial gauge, $\delta g_{z \mu} =  0$ and $\delta A_{z} = 0$, we must consider the full set of perturbations
\be
\delta A_t \; ,\; \delta g_{tt} \; ,\; \delta g_{xx} \; ,\; \delta g_{yy}
\nn\ee
The equations governing these perturbations were derived in \cite{edalati1, edalati2}, although the static susceptibility was not calculated. We relegate details of this calculation to the Appendix. In brief, it proceeds by first eliminating $\delta g_{xx}$ to give three coupled, ordinary differential equations  for the static perturbations $ \delta A_t$,  $\delta g_{tt}$ and $\delta g_{yy}$. We  solve these numerically and extract the static susceptibility $\chi(k)$.
%
%
%
%
%
%



\begin{figure}
\begin{center}
\includegraphics[width=2.6in]{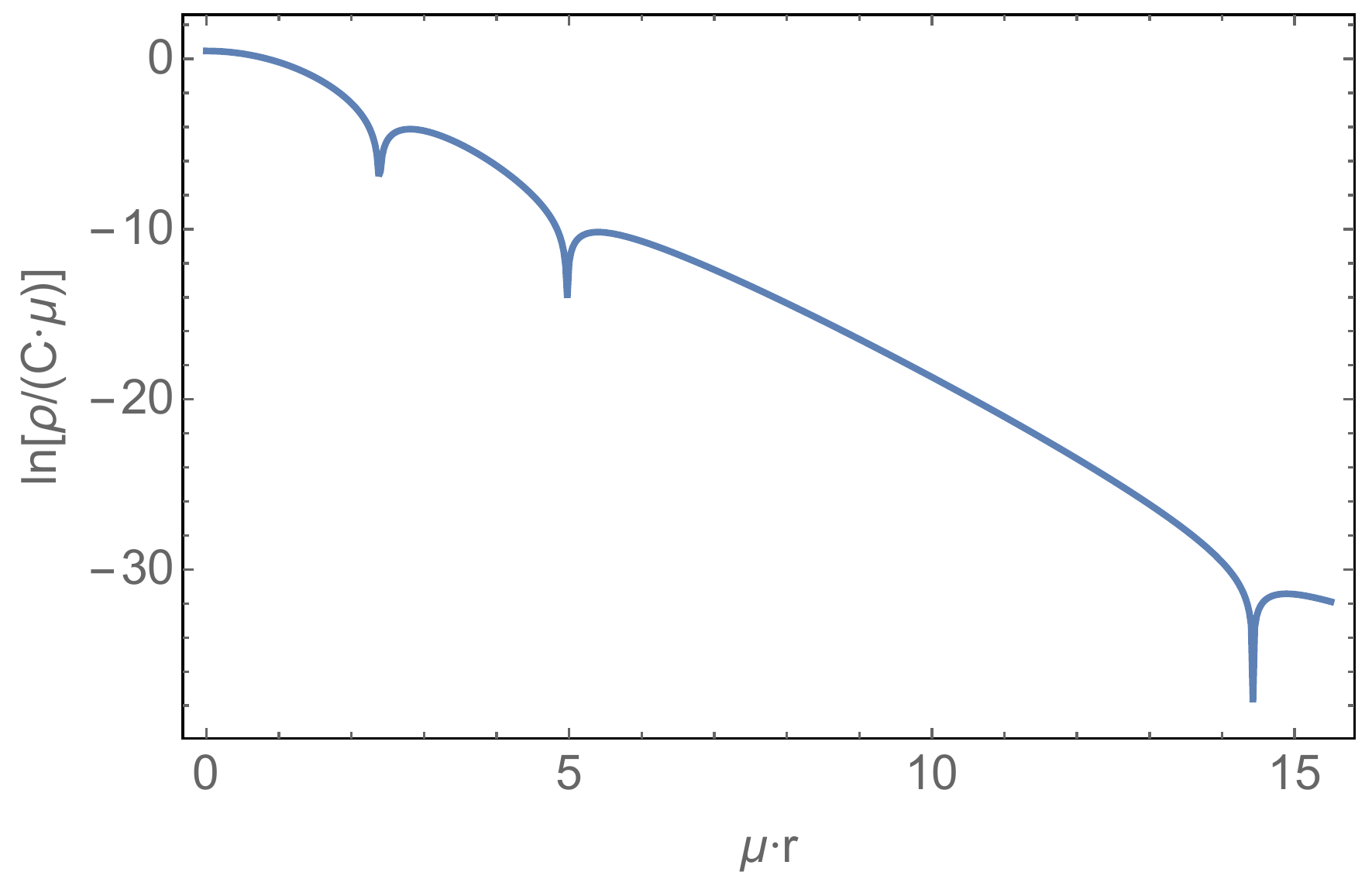}\ \ \ \  
\caption{$T<T_c$, $\mu_0\neq 0$. The induced charge density oscillates at large distances, shown here for an impurity of width $R\mu_0=1$ and temperature $T/\mu_0 = 0.2$. }
\label{oscillations}
\end{center}
\end{figure}

\para Armed with susceptibility, we can perform the Fourier transform and calculate the induced charge density. For high temperatures, $T\gg \mu_0$, the response  in charge density is qualitatively similar to that of the Schwarzchild metric. In particular, the system exhibits exponential Debye-like screening \eqn{goeslikethis} at large distances. However, this behaviour changes below a critical temperature which, numerically, we find to be
\be T_c\approx 0.33\mu_0\nn\ee
For $T<T_c$, the induced charge density oscillates. At long distances, the charge density is given by $\rho(r) = \rho_0 + \delta\rho(r)$ where $\rho_0=\mu_0/z_+$ is the background charge, while $\delta \rho$ takes the form
\be \delta \rho(r) \sim  \frac{e^{-r/\lambda}}{\sqrt{r}}\,\cos(r/\xi)\label{decay}\ee
with the length scales $\lambda$ and $\xi$ set by $\mu_0$ and $T$. This charge density is shown in Figure \ref{oscillations}. As before,  nodes in the charge density appear as cusps in $\log(|\rho|/C\mu)$. The appearance of multiple cusps shows that the charge density is oscillating.

%
 
\begin{figure}
\begin{center}
\includegraphics[width=2.3in]{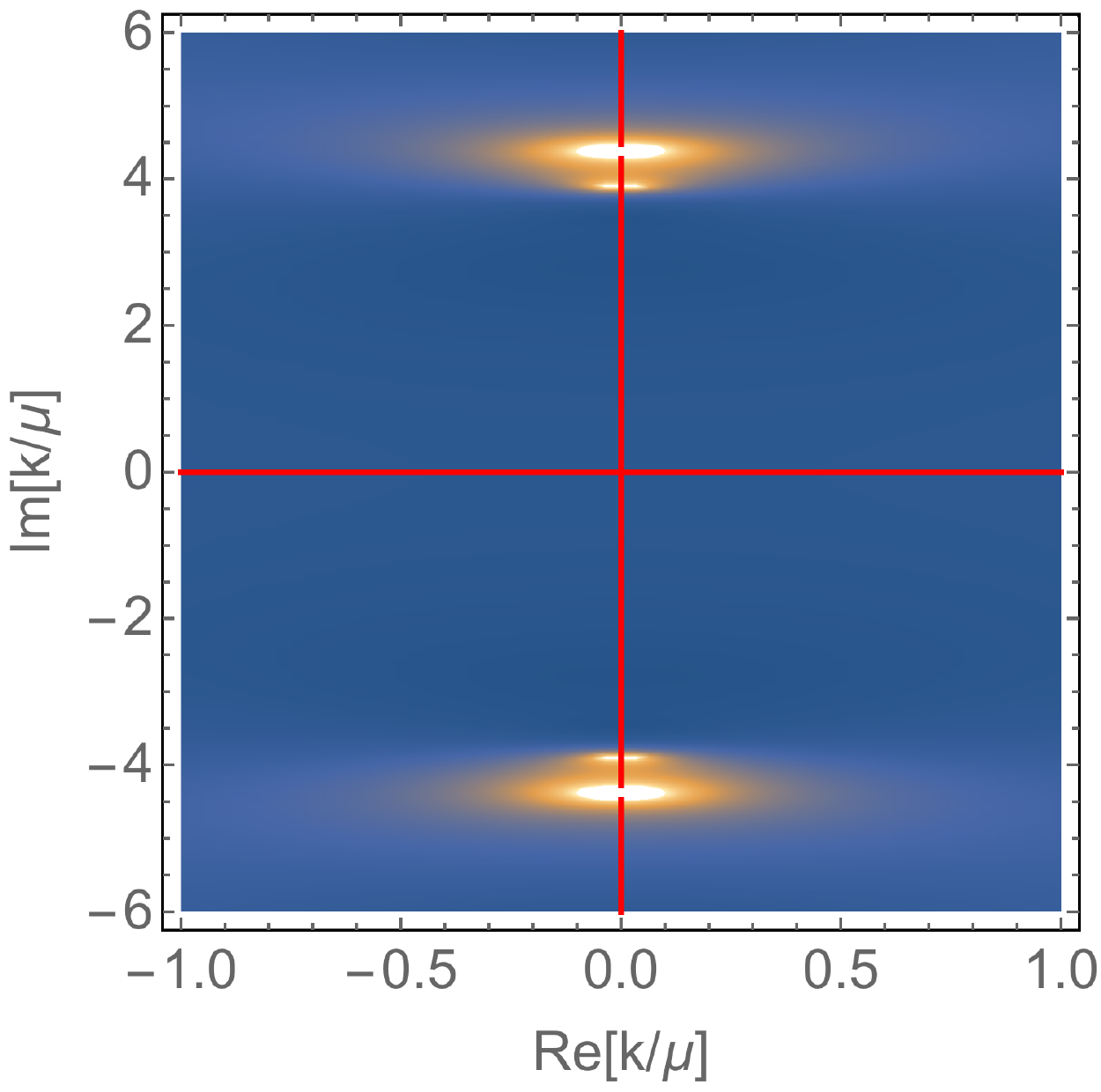}\ \ \ \ 
\includegraphics[width=2.3in]{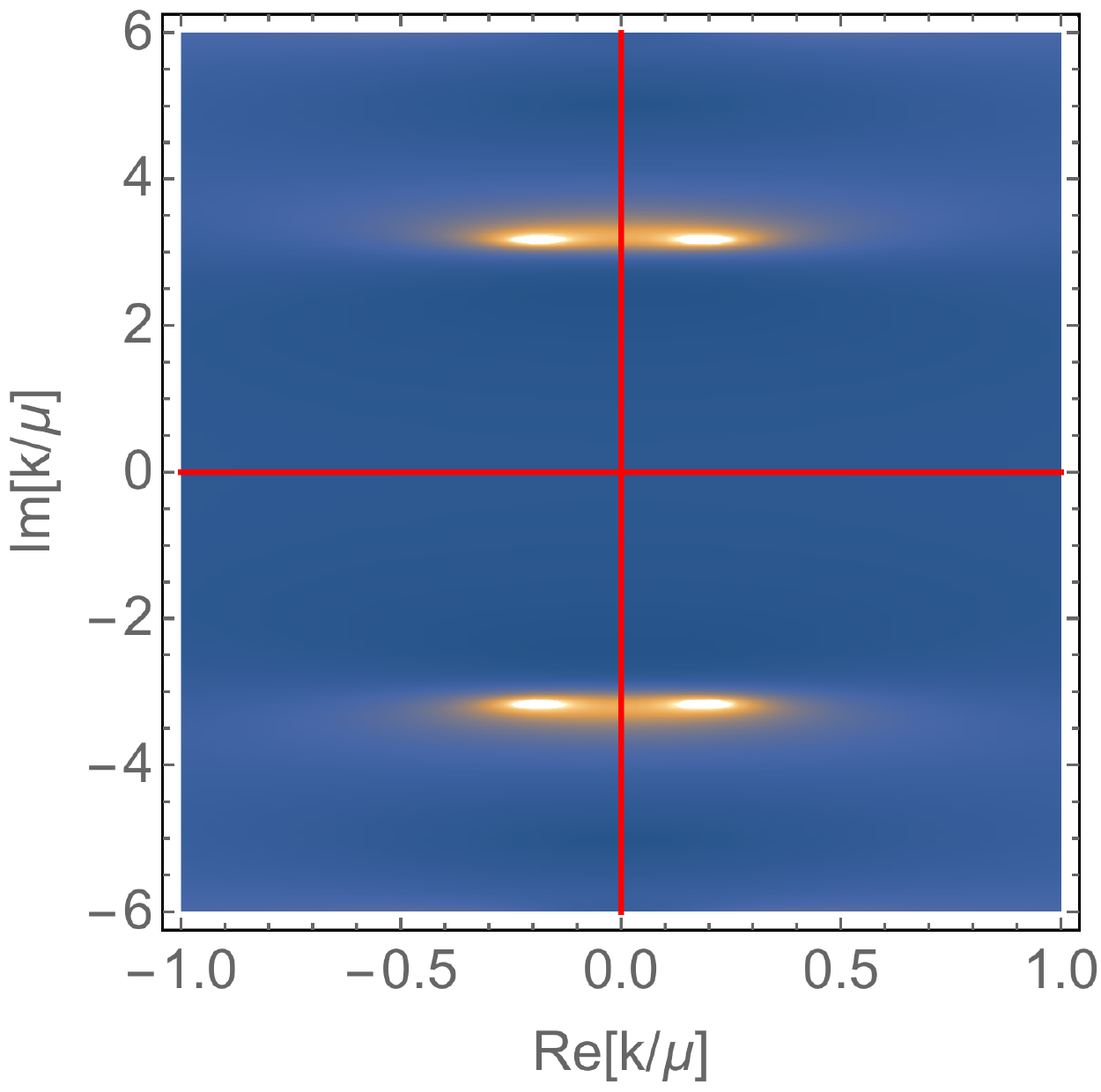}\ \ \ \ 
\caption{$T\approx T_c$, $\mu_0\neq 0$: The absolute value of the density-density correlator in the complex momentum plane above the phase transition at $T=0.39\mu_0>T_c$ (left) and below the phase transition at $T = 0.295\mu_0<T_c$ (right).}
\label{figdensity}
\end{center}
\end{figure}

 \para
  The origin of these oscillations can again be understood by looking  in the complex momentum  plane. At temperatures $T\geq T_c$, the susceptibility exhibits a string of poles along the imaginary axis, just as we saw for the Schwarzchild black hole. This is the situation depicted in the left of Figure \ref{figdensity} where two of the poles are visible. This results in the now-familiar exponential damping of Debye screening of the form \eqn{goeslikethis}.  

\para As we lower the temperature, the two poles depicted in the figure get closer together. Eventually, at the critical temperature $T=T_c$, these two poles merge.  As the temperature is lowered yet further, the poles move off the imaginary axis and into the complex momentum plane, gaining a real part. This can be seen in right hand plot of Figure \ref{figdensity}. 

\para
The observed oscillations in the induced charge density for $T<T_c$ can be traced to the position of the pole away from the imaginary axis. 
The Fourier transform is dominated by the pole, with the  $e^{i{\bf k}\cdot {\bf r}}$ factor now contributing an oscillatory piece. If we write the location of the pole as  $k_\star=k_1(T)+ik_2(T)$  then, for $T<T_c$, the  long-distance profile of charge density takes the form \eqn{decay} with $\lambda=1/k_2$ and $\xi=1/k_1$.
%
%


  \para
 Charge screening of the form \eqn{decay} is very similar to that seen in Friedel oscillations at finite temperature, where the damping factor takes the form $e^{-Tr}$. The difference between the two is only quantitative:  for $T\ll \mu$, the Friedel oscillations are very weakly damped while, for the black hole, the real part of $k_\star$ is always comparable to the imaginary part as we lower the temperature.  This will ultimately lead to a more dramatic  difference between the oscillations identified above and Friedel oscillations at zero temperature.

\begin{wrapfigure}{r}{0.37\textwidth}
\label{figtraj}
\includegraphics[width=0.37\textwidth]{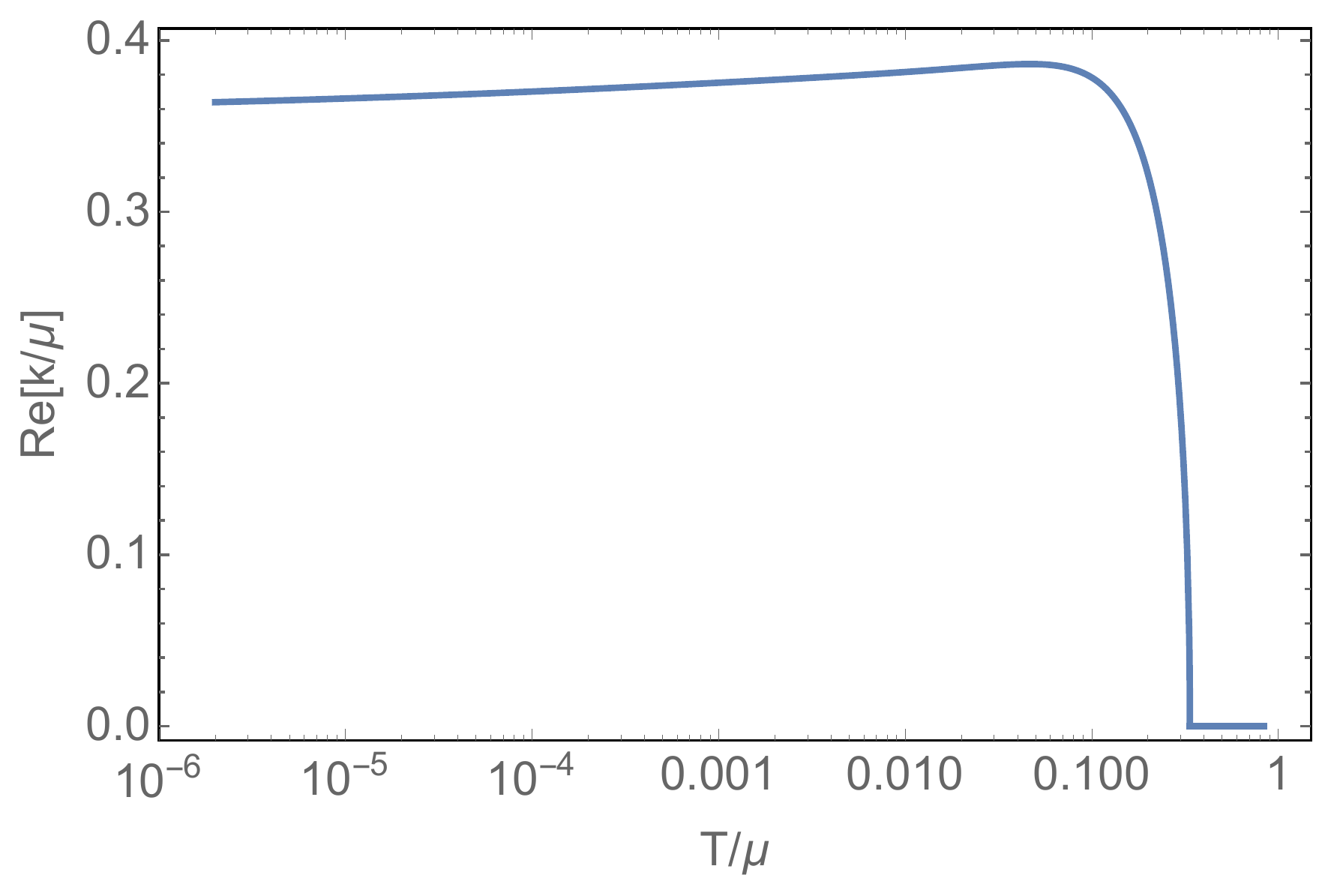}
\caption{The trajectory traced by the real part of the pole as the temperature is lowered.} 
\end{wrapfigure}

\para This qualitative change in the position of the pole at $T=T_c$ can be viewed as a kind of second order phase transition, with ${\rm Re}(k)$ playing the role of the order parameter. (We stress, however, that this is not a real phase transition in the sense that  $k=0$ thermodynamic quantities remain analytic). This is clearly seen if plot the trajectory of ${\rm Re}(k)$ for the lowest pole as we vary the temperature. (Note that, for $T > T_c$, the trajectory is actually two poles on the imaginary axis). Numerically we find that, close to the transition point, the behaviour of ${\rm Re}(k)$ is well modelled by the mean-field exponent,
 \be {\rm Re}(k) \sim (T_c-T)^{1/2}\ \ \ \ \ \ \ \ T<T_c\nn\ee
A similar second order phase transition was observed previously in a phenomenological model of QCD, involving nucleons interacting with pions  at finite density \cite{sivak1}. However, the phase transition appears not to occur in weakly coupled non-Abelian gauge theories where poles in complex plane are observed at finite density \cite{sivak2}, but their effects are washed out at long distance by the more familiar Friedel oscillations.

\subsubsection*{Back to Zero Temperature}

\begin{figure}
\begin{center}
\includegraphics[width=2.3in]{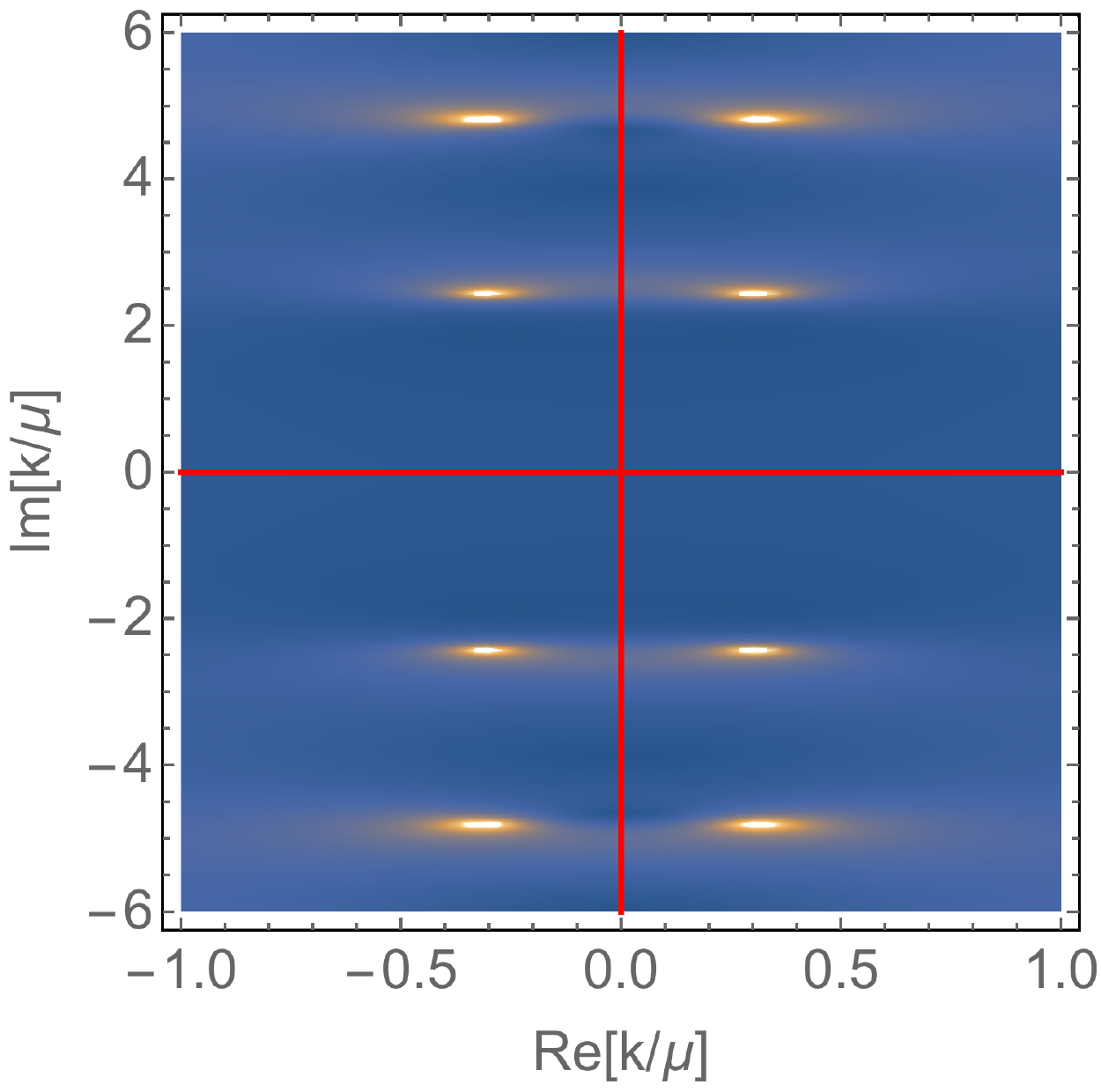}\ \ \ \  
\includegraphics[width=2.3in]{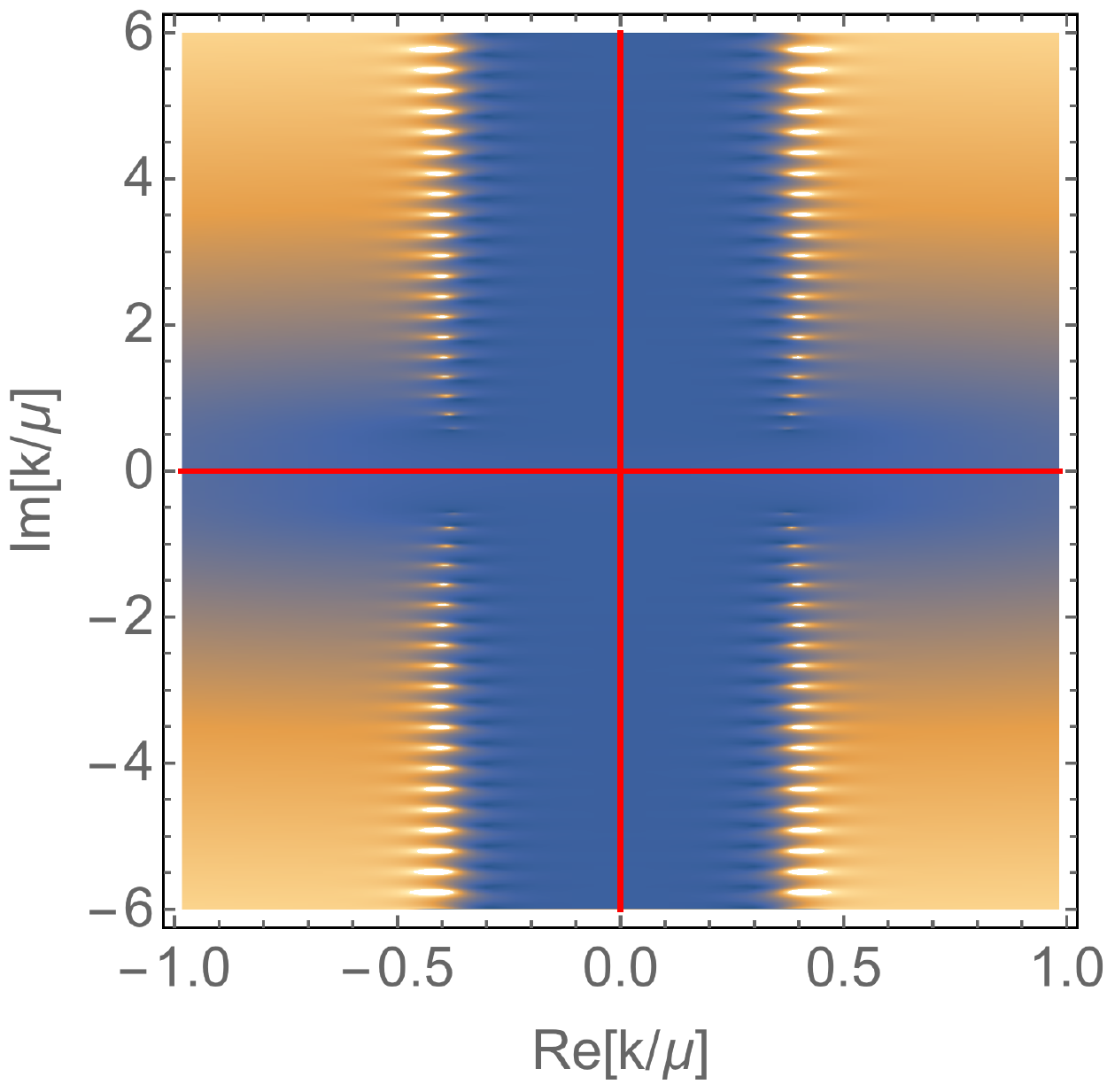}
\caption{$T<T_c$, $\mu_0\neq 0$: The absolute value of the density-density correlator in the complex momentum plane, for $T/\mu_0 = 0.21$ (left) and $T/\mu_0=0.0006$ (right).}
\label{figstring}
\end{center}
\end{figure}

Finally, we can ask: what becomes of these oscillations as we approach $T\rightarrow 0$ and the \RN black hole becomes extremal? We find that the solitary pole that we've been following asymptotes to a fixed location in the complex plane, away from both real and imaginary axes. Moreover, as the temperature is lowered, this solitary pole is joined by a string of others. This is shown in Figure \ref{figstring}. Each of these poles left the safety of the imaginary axis at some temperature, typically one larger than $T_c$. (Indeed, for certain regimes of the parameters, we find that these higher poles can lead to one or two anomalous oscillations even at $T>T_c$ which occur at finite $r$ before they become suppressed at large $r$). 

\para
Ultimately, it appears that these poles coalesce to form a branch cut. 
It is worth mentioning that the existence of a branch cut, lying roughly parallel to the imaginary momentum axis, is  reminiscent of the situation in Fermi liquids at zero temperature. There the branch cut extends down to the real axis which ensures the power-law fall-off of Friedel oscillations. A number of studies of $2k_F$ singularities in various non-Fermi liquid states find that the end of the branch-cut  remains at a real value of the momentum \cite{nayak,altshuler,mross,gonzalo}.

\begin{figure}
\begin{center}
\includegraphics[width=2.6in]{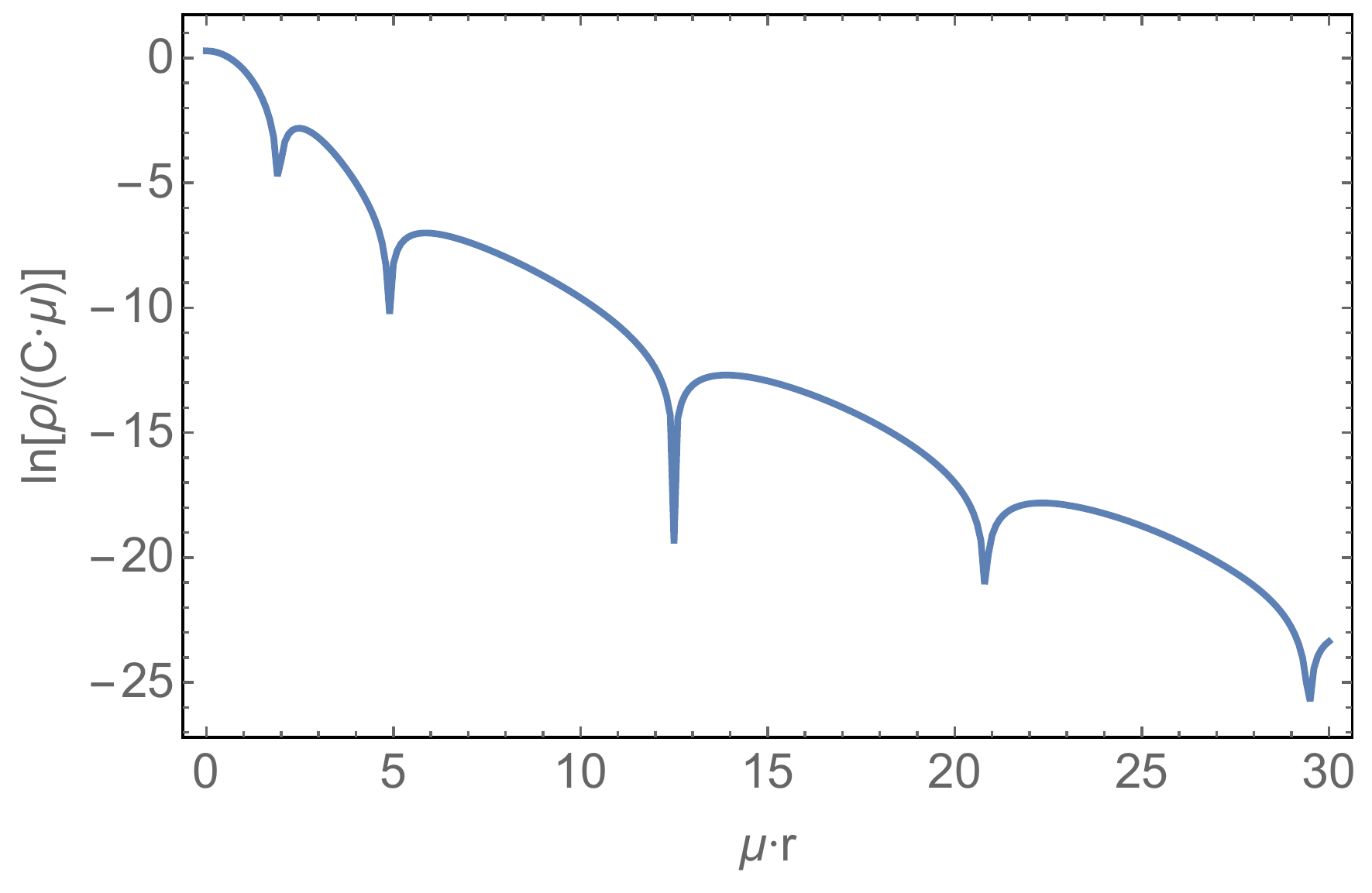}\ \ \ \  
\includegraphics[width=2.6in]{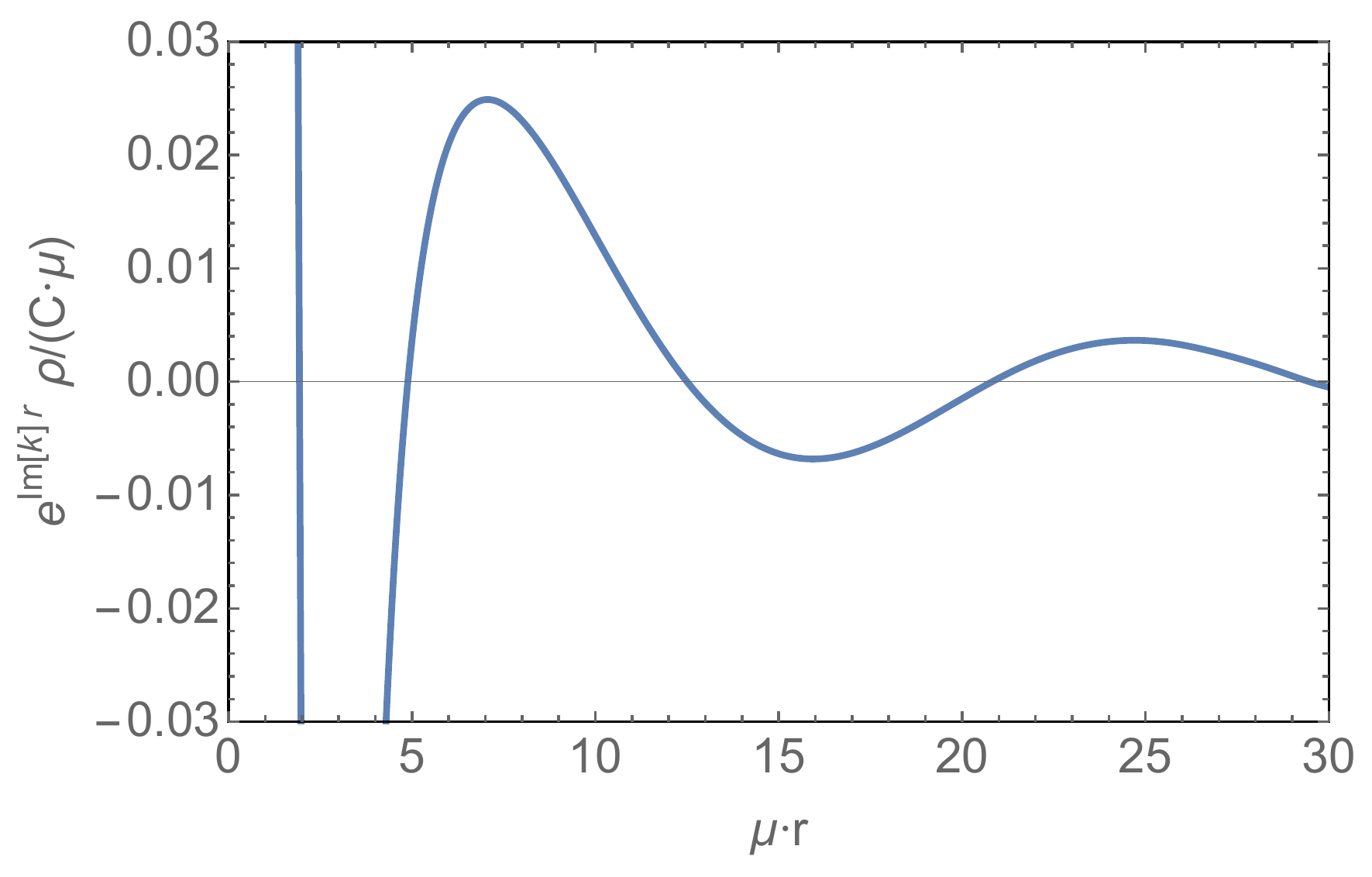}
\caption{$T=0$. The zero temperature response gives a series of oscillations (left). These can be seen more clearly by factoring out the exponential damping (right). Both plots use an impurity of width $R\mu_0=1$.}
\label{zeroToscillation}
\end{center}
\end{figure}

\para
In contrast, in our holographic model, the branch cut terminates in the complex plane and, consequently, the screening at zero temperature is not greatly changed from \eqn{decay}: the exponential suppression and oscillations both remain although, in  principle, the accompanying power-law may change depending on the cut residue. This $T=0$ screening is plotted in Figure \ref{zeroToscillation}. The left-hand plot shows $\log(\rho/C\mu_0)$, with the nodes appearing as cusps. To exhibit the oscillations more clearly, in the right-hand plot we have rescaled the charge density by the exponential suppression factor $e^{i{\rm Im}(k_\star )r}$, with $k_\star$ the end of the branch cut.

%

\begin{figure}
\begin{center}
\includegraphics[width=2.6in]{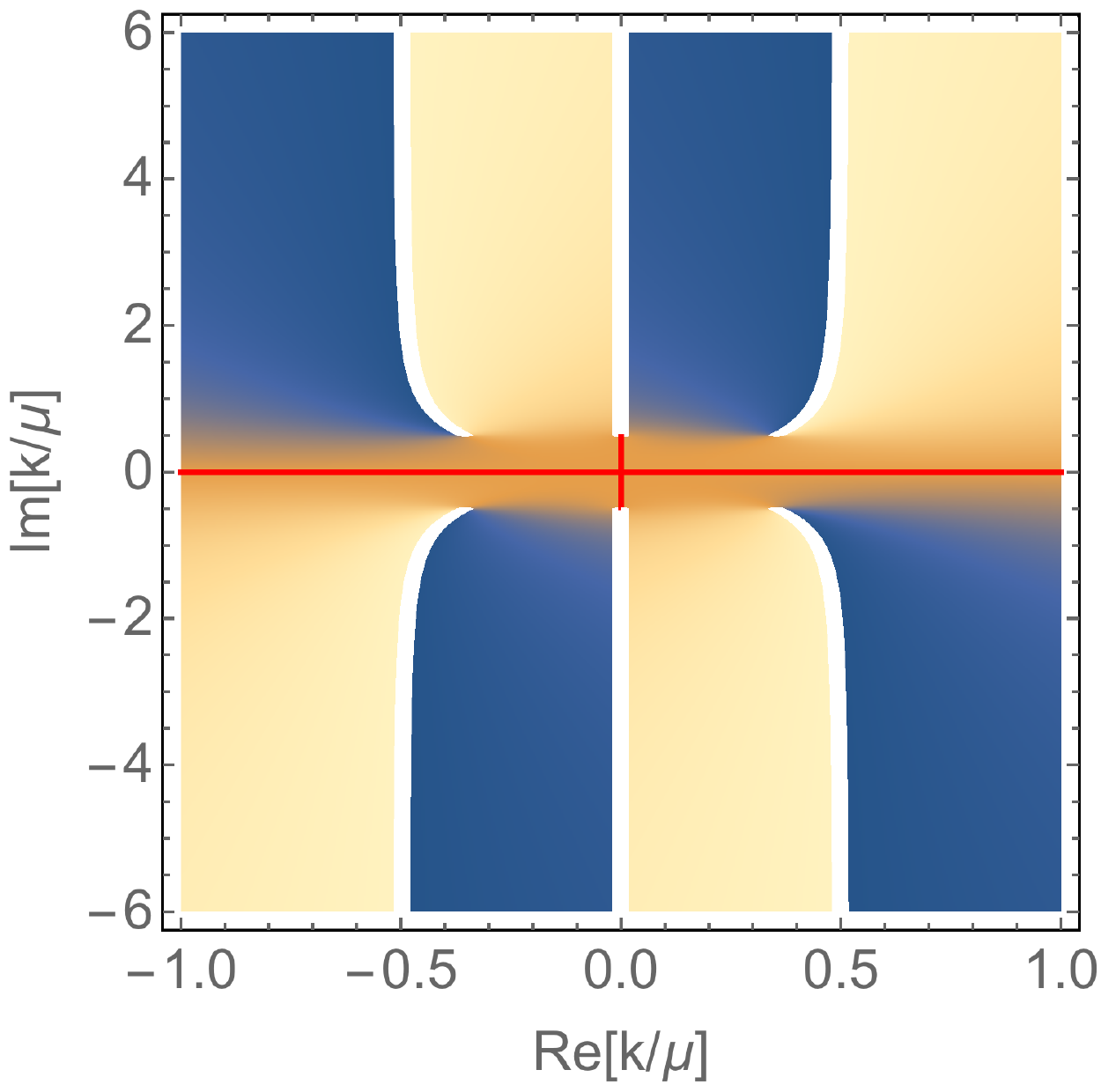}
\caption{$T=0$, $\mu_0\neq 0$: The argument of $\nu_-$ in the complex momentum plane. The branch cuts  along ${\rm Re}(\nu_-)=0$ should be compared to the string of poles shown in Figure 7.  }
\label{figbranch}
\end{center}
\end{figure}

\para
The existence of this branch cut  at zero temperature should be viewed as the underlying cause  of our charge oscillations. 
 There is compelling reason to believe that this branch cut is associated to the AdS$_2\times {\bf R}^2$ near horizon region of the geometry. This is the regime of local criticality, which can be thought of as a scale invariant theory with dynamical exponent $z\rightarrow \infty$, so that time and energies scale, while space and momenta do not. This means that the current operators in the theory are labelled by their momentum $k$ and, in the far infra-red, have dimension $\delta_\pm(k) = \frac{1}{2} + \nu_\pm(k)$, where
\be\label{eq:nu_def} \nu_\pm(k) = \frac{1}{2}\sqrt{5+8\left(\frac{k}{\mu_0}\right)^2 \pm 4 \sqrt{1+ 4\left(\frac{k}{\mu_0}\right)^2}}\ee
In this paper we are interested in complex momenta. As we take the limit $T\rightarrow 0$, we observe that the poles converge to a curve which is very well approximated by the branch cut defined by
\be {\rm Re}(\nu_-)=0\label{numinus}\ee
In particular, the location of the end point of the branch point lies at $k/\mu_0=1/2\sqrt{2} + i/2$ and  this appears to be the final resting place of the leading pole. 
In Figure \ref{figbranch}, we plot the argument of $\nu_-$. The branch cuts defined by ${\rm Re}(\nu_-)=0$ lie in the complex plane. There are also additional branch cuts along the imaginary axes, associated to the embedded square-root in \eqn{eq:nu_def}, which do not correspond to poles in the \RN background.

\para
The agreement of the AdS$_2$ and \RN branch cuts requires an explanation. It is certainly true that AdS$_2$ correlation functions (at finite $T$ or finite $\omega$) will exhibit such a branch cut in the complex momentum plane. Moreover, it is well known that the AdS$_2$ region dominates the spectral density of the theory at low-frequencies \cite{edalati2,seandiego} and this can be understood using the kind of matching procedure pioneered in \cite{john} between near- and far-horizon geometries. (See, for example, the appendix of \cite{aristossean} for a review of this procedure). However, it is unclear to us how to implement this procedure in the present context.

%
%

 \para The fact that the accumulation of poles appears to converge towards the locus \eqn{numinus} is strong evidence that the oscillations are due to local criticality and there is an intuitive way to  understand this behaviour. The near-horizon region of the geometry supports  low-energy excitations with a range of momentum $k<\mu_0$ and therefore a minimum wavelength which is roughly $\sim 1/\mu_0$. As we reviewed in the introduction, the essence of Friedel oscillations is that the modes which do the screening have a finite size. Here too, we see that modes have a finite size, now with a range of wavelengths giving rise to the exponential behaviour rather than power-law. 
The lesson is similar to that in \cite{seandiego}: low-energy modes at finite momentum occur in both Fermi liquids and in locally critical theories, ensuring that the two states share certain features. In \cite{seandiego}, the focus was on efficient umklapp scattering. Here we see that charge oscillations due to the screening of an impurity can be added to this list.


\para
We finish with some speculation. It is natural to wonder if, in other models, with more bells and whistles, one could move the branch cut so that  it terminates on the real momentum axis. This would then result in $T=0$, Friedel-like oscillations, with power-law decay. However, the fact that the branch cut lies at ${\rm Re}(\nu_-)=0$ means that  a new phenomenon   accompanies the power-law oscillations: the onset of a finite wavelength instability. Such instabilities have been seen previously in a number of different holographic models \cite{hirosi,ja1,ja2}. In all of these cases, the infra-red dimension $\delta_\pm(k)$ of
some operator violates the BF bound for some finite wavenumber which, at the onset of the
instability,  translates into the requirement that $\nu_\pm(k)=0$ for some real
momentum $k$. This suggests that, within holography, the emergence of Friedel oscillations from the bosonic geometry would indicate that the system lives on the edge of an instability.  It would be interesting to explore this connection
further.

\section*{Acknowledgements}

We're grateful to David Berman, Jan de Boer, Sean Hartnoll, Gary Horowitz, 
Nabil Iqbal, Jorge Santos, Gonzalo Torroba and Benson Way for some 
combination of comment, discussion, correspondence, and sharing of 
\cite{jorge}.
 We are supported by STFC and by the European Research Council under the European Union's Seventh Framework Programme (FP7/2007-2013), ERC Grant agreement STG 279943, Strongly Coupled Systems. MB is funded by Churchill College.

\appendix

\section{Appendix}

In this Appendix we discuss in detail the calculation of the static susceptibility in the \RN background. We perturb the solution by introducing $\delta A_t(z,x) = \delta A_t(z) e^{i k x}$. As we discussed in the main text we will work in radial gauge
\be
\delta g_{z \mu} = 0 \;\;\;\;\;\;\;\; \delta A_{z} = 0\nn
\ee
and so the other fields we must turn on are $\delta g_{xx}$, $\delta g_{yy}$ and $\delta g_{tt}$. The perturbation equations take a simpler form if we raise one index on these fields using the background metric. Then the linearised Maxwell equation reads
\be
 f(z) \delta A_t'' - k^2 \delta A_t + \frac{\mu_0 f(z)} {2 z_+}  \bigg( (\delta g^t_{\;\;t})' - (\delta g^x_{\;\;x})' - (\delta g^y_{\;\;y})' \bigg) = 0 \label{at}
\ee
while we also have a host of Einstein equations
\begin{eqnarray}
f(z) (\delta g^{y}_{\;\;y})'' + f'(z) (\delta g^{y}_{\;\;y})' - \frac{3 f(z)}{z}( \delta g^y_{\;\;y})' - \frac{f(z)}{z}\bigg( (\delta g^x_{\;\;x})' + (\delta g^t_{\;\;t})' \bigg) && \label{gyy} \\ - \frac{\mu_0 z^2}{z_+}  \delta A_t' - k^2 \delta g^y_{\;\;y}  - \frac{ \mu_0^2 z^2}{2 z_+^2} \delta g^t_{\;\;t} &=0& \nn \;\;\;\;\; \\
f(z) (\delta g^{x}_{\;\;x})'' + f'(z)(\delta g^x_{\;\;x})' - \frac{3 f(z)}{z} (\delta g^x_{\;\;x})' - \frac{f(z)}{z}\bigg( (\delta g^y_{\;\;y})' + (\delta g^t_{\;\;t})' \bigg) && \label{gxx} \\ - \frac{\mu_0 z^2}{z_+} \delta A_t' - k^2 \delta g^y_{\;\;y}  - \bigg( k^2 + \frac{\mu_0^2 z^2}{2 z_+^2} \bigg)\delta g^t_{\;\;t}  &=&0  \nonumber \\
f(z)  (\delta g^{t}_{\;\;t})'' - \frac{3 f(z)}{z} (\delta g^t_{\;\;t})' + \frac{3 f'(z)}{2} (\delta g^t_{\;\;t})' + \bigg (\frac{f'(z)}{2} - \frac{f(z)}{z}\bigg)\bigg((\delta g^x_{\;\;x})' + (\delta g^y_{\;\;y})'\bigg) \label{gtt} \\
+ \frac{ \mu_0 z^2}{z_+} \delta A_t' + \bigg (\frac{ \mu_0^2 z^2}{2 z_+^2} - k^2 \bigg) \delta g^t_{\;\;t}&=0&\nonumber \\ 
f(z) \bigg( (\delta g^{t}_{\;\; t})'' + (\delta g^x_{\;\;x})'' +  (\delta g^y_{\;\;y})'' \bigg) - \frac{f(z)}{z} \bigg( (\delta g^{t}_{\;\;t})' + (\delta g^x_{\;\;x})' + (\delta g^y_{\;\;y})' \bigg) \label{allg} \\ + \frac{f'(z)}{2} \bigg( 3 (\delta g^{t}_{\;\;t})' + (\delta g^x_{\;\;x})' + (\delta g^y_{\;\;y})' \bigg)  
+ \frac{ \mu_0 z^2}{z_+} \delta A_t' + \frac{ \mu_0^2 z^2} {2 z_+^2} \delta g^t_{\;\;t}  &=0 & \nonumber \\ 
2 f(z) (\delta g^t_{\;\;t})' + f'(z) \delta g^t_{\;\;t} + 2 f(z) (\delta g^y_{\;\;y})' + \frac{2 \mu_0 z^2}{z_+} \delta A_t =0&&  \label{constraint}
\end{eqnarray}
Whilst this is a complicated system of equations, one can simplify the calculation of the susceptibility by finding a subsystem of equations that govern the fluctuations of $\delta A_x, \delta g^t_{\;\;t}, \delta g^y_{\;\;y}$. This is possible because the Einstein equations imply an algebraic expression for $(\delta g^x_{\;\;x})'$ in terms of the other fields and their derivatives. To see this one takes the linear combination of equations  \eqn{allg} $-$ \eqn{gtt} $-$ \eqn{gxx} $-$ \eqn{gyy}. This yields the relation
\be
(\delta g^x_{\;\;x})' &=& \frac{z}{4f(z) - z f'(z)} \bigg[ f'(z) (\delta g^y_{\;\;y})' - \frac{4f(z)}{z}\bigg( (\delta g^y_{\;\;y})' + (\delta g^t_{\;\;t})' \bigg) \label{hxxprime} \\ && \ \ \ \ \ \ \ \ \ \ \ \ \ \ \ \ \ \ \ \ \ \ \ \ - \frac{2 \mu_0 z^2}{z_+} \delta A_t'  - 2k^2 \delta g^y_{\;\;y} - \bigg(2k^2 + \frac{\mu_0^2 z^2}{z_+^2} \bigg) \delta g^t_{\;\;t} \nonumber \bigg ]
\ee
We can then construct a closed set of ordinary differential equations for $\delta A_t$, $\delta g^y_{\;\;y}$ and $\delta g^t_{\;\;t}$ by using this identity to eliminate $(\delta g^x_{\;\;x})'$ from equations \eqn{at}, \eqn{gyy} and \eqn{constraint}. The resulting equations are first order in $\delta g^t_{\;\;t}$ and second order in the fields $\delta A_t$ and $\delta g^y_{\;\;y}$. This means that we  need to fix five constants of integration to obtain a unique solution. 

\paragraph{}At the horizon, we demand regular behaviour in general. At finite temperature this is achieved by imposing the analytic expansion
\begin{eqnarray}
\delta A_t &\sim& c_1(z - z_+) + \mathcal{O}((z-z_{+})^{2}) \nonumber \\ 
\delta g^y_{\;\;y} &\sim& c_2+ \mathcal{O}(z-z_{+})  \label{boundarycond1}\\
\delta g^t_{\;\;t} &\sim& c_3(z - z_+)+ \mathcal{O}((z-z_{+})^{2})\nn
\end{eqnarray}
where $c_1$, $c_2$ and $c_3$ are constants. Plugging this expansion in the equations of motion fixes $c_{3}$ as a linear combination of $c_{1}$ and $c_{2}$. Imposing these boundary conditions removes the three modes that lead to singular behaviour near the horizon. Of the five original modes, we are therefore left with two, corresponding to the unconstrained constants of integration $c_{1}$ and $c_{2}$. 

\paragraph{}At zero temperature, which happens when $\mu_0\,z^{(0)}_{+}=\sqrt{12}$, the singular point at $z=z^{(0)}_{+}$ is irregular and we impose the power series expansion
\begin{eqnarray}
\delta A_t &\sim& c^{+} \,(z^{(0)}_{+}-z)^{3/2+\nu_{+}(k/\mu_0)}\,(1+\mathcal{O}(z^{(0)}_{+}-z )) \nonumber \\ 
&+&\,c^{-}\,(z^{(0)}_{+}-z)^{3/2+\nu_{-}(k/\mu_0)}\,(1+\mathcal{O}(z^{(0)}_{+}-z ))  \nonumber \\ 
\delta g^t_{\;\;t} &\sim & c^{+} d^{+}_1(k/\mu_0)\,(z^{(0)}_{+}-z)^{3/2+\nu_{+}(k/\mu_0)}\,(1+\mathcal{O}(z^{(0)}_{+}-z ))\nonumber \\ 
&+&c^{-}\,d^{-}_1(k/\mu_0)\,(z^{(0)}_{+}-z)^{3/2+\nu_{-}(k/\mu_0)}\,(1+\mathcal{O}(z^{(0)}_{+}-z )) \label{boundarycond2}\\ 
\delta g^y_{\;\;y} &\sim&c^{+} d^{+}_2(k/\mu_0)\,(z^{(0)}_{+}-z)^{1/2+\nu_{+}(k/\mu_0)}\,(1+\mathcal{O}(z^{(0)}_{+}-z ))\nonumber \\ 
&+&c^{-}\,d^{-}_2(k/\mu_0)\,(z^{(0)}_{+}-z)^{1/2+\nu_{-}(k/\mu_0)}\,(1+\mathcal{O}(z^{(0)}_{+}-z ))
\nn
\end{eqnarray}
with $\nu_{\pm}$ as in equation \eqn{eq:nu_def}. The constants $d_{i}^{\pm}$ are determined by the indicial system of equations for the modes and are functions of  $k/\mu_0$, while the $c^{\pm}$ are the two unconstrained modes near the horizon.

\paragraph{}In the UV, the asymptotic expansion of the modes is fixed in terms of the constants of integration $\left\{ s_{i},\,v_{j}\right\}$ 
\begin{eqnarray}\label{eq:UV_exp}
 \delta A_t &\rightarrow& s_{1}+v_{1}\,z+\cdots  \nonumber \\
  \delta g^y_{\;\;y} &\rightarrow& s_{2}+\cdots+v_{2}\,z^{3}+\cdots  \label{naivedata} \\
 \delta g^t_{\;\;t} &\rightarrow& s_{3}+\cdots .\nn
 \end{eqnarray}
As expected, $ \delta g^t_{\;\;t} $, which satisfies a first order equation, gives only one constant of integration close to the boundary. Proceeding naively, one might wish to set $s_{2}=s_{3}=0$ since these correspond to sources of the stress tensor of the boundary theory. At the same time we would like to impose $s_{1}=1$ which would then correspond to a source for the charge density. However, it is clear we do not have enough freedom to apply all of the conditions. We have already fixed three constants of integration by our regularity conditions, \eqn{boundarycond1} or \eqn{boundarycond2}, and so can only fix two more in the UV. 

\paragraph{}This issue is a consequence of the fact we are working in radial gauge (an analogous situation was found in holographic conductivity calculations in \cite{Qlattices}). The gauge transformation which sets $\delta g_{z \mu} = 0$ is a large gauge transformation that acts on the boundary data. One can therefore only apply boundary conditions that are gauge equivalent to $s_{2}=s_{3}=0$. That is, we impose
\begin{eqnarray}
\delta A_t &\rightarrow& s_{1} \nonumber \\
\delta g^t_{\;\;t} - \delta g^y_{\;\;y}  &\rightarrow& 0 \nn
\end{eqnarray}
Solving our system of equations with these boundary conditions produces a unique solution satisfying $s_2 = s_3=s$. Matching the solutions from the two boundaries now will completely fix the remaining five constants $\left\{c_{i=1,2},\,v_{i=1,2},\,s\right\}$ or $\left\{c^{\pm},\,v_{i=1,2},\,s\right\}$. This solution can then be mapped to one satisfying $s_{2}=s_{3}=0$ by performing a gauge transformation of the form $x \rightarrow x + f(x,z), z \rightarrow z + h(x,z)$ such that
\be
f(x,z) &\rightarrow& \frac{s_{3}}{2 i k} e^{ i k x}\, +\mathcal{O}(z)\nn\\
h(x,z) &\rightarrow& \frac{s_{3}}{2} e^{ i k x}\,z+\mathcal{O}(z^{2})\nn
\ee
close to $z=0$ and are smooth close to the horizon. Of course, performing this gauge transformation turns on non-zero perturbations in $g_{zz}$ and $g_{zx}$ and so breaks the radial gauge condition.

\paragraph{}
Nevertheless, after performing this coordinate transformation, the metric dies off suitably fast at the boundary. In these coordinates, the static susceptibility is therefore given by the radial derivative of $\delta A_t$ at the boundary. Using our expansion \eqn{eq:UV_exp} we have
\be
\chi(k) = -  \frac{v_{1}}{s_1}+\frac{\mu_0\,s_{3} }{2 z_{+} \,s_1} \nn
\ee


\end{document}